\documentclass[prd,aps,11pt]{revtex4}

\usepackage{epsf}
\renewcommand{\email}[1]{\tt #1}

\begin{document}
\vspace*{1.cm}
\title{Is it possible to consider Dark Energy and Dark Matter as a same and unique Dark Fluid?\vspace*{1.cm}}

\author{Alexandre Arbey\footnote{E--mail: \email{arbey@obs.univ-lyon1.fr}}}
\vspace*{1.cm}
\affiliation{Centre de Recherche Astronomique de Lyon (CRAL), 9 avenue Charles Andr\'e, 69561 Saint Genis Laval Cedex, France\vspace*{1.cm}}
\begin{abstract}
\vskip 0.5cm
\noindent In the standard model of cosmology, the present evolution of the Universe is determined by the presence of two components of unknown nature. One of them is referenced as ``dark matter'' to justify the fact that it behaves cosmologically like usual baryonic matter, whereas the other one is called ``dark energy'', which is a component with a negative pressure. As the nature of both dark components remains unknown, it is interesting to consider other models. In particular, it seems that the cosmological observations can also be understood for a Universe which does not contain two fluids of unknown nature, but only one fluid with other properties. To arrive to this conclusion, we will review the observational constraints from supernov\ae~of type Ia, cosmic microwave background, large scale structures, and the theoretical results of big-bang nucleosynthesis. We will try to determine constraints on this unifying ``dark fluid'', and briefly review different possibilities to build models of dark fluid.\vspace*{2.cm}
\end{abstract}

\vskip 1.cm
\maketitle

\noindent In the standard model of cosmology, the total energy density of the Universe is dominated today by the densities of two components: the first one, called ``dark matter'', is generally modeled as a system of collisionless particles (``cold dark matter'') and consequently has an attractive gravitational effect like usual matter. The second one, generally refered as ``dark energy'' or ``cosmological constant'' can be considered as a vacuum energy with a negative pressure, which seems constant today. The real nature of these two components remains unknown. The reason of such a distinction between both dark components is mainly historical. Indeed, the cosmological constant was introduced by Einstein in order to justify the existence of a static Universe. Because this constant can appear naturally in the Einstein equations as a new fundamental constant, introducing it was not problematic. Moreover, no distinction was made between the dark matter and the usual baryonic matter. Therefore, at this epoch, the standard model of cosmology was not so different from the present one, with however an interesting conceptual difference: the two components present in the cosmological models were at that moment not considered of unknown nature, because, firstly no dark matter was needed, and secondly the cosmological constant could be considered as a fundamental constant of Nature. Thus, the Universe was finally not so strange at that time, and the distinction between cosmological constant and matter still holds. Later on important problems have appeared. First, big--bang nucleosynthesis has shown that the baryons represent only a very small part of the matter density of the Universe \cite{introBBN}, so that dark matter has to exist and to represent a large fraction of the total density of the Universe. Moreover, this dark matter has to be non--baryonic, and yet the standard model of particle physics does not provide good candidates for such a kind of matter. No particle of dark matter have been detected yet, and the nature of the dark matter remains uncertain. Usual candidates for dark matter are based on new particle physics theories, like supersymmetry, but as these new theories have not been verified yet and as some simulations based on cold dark matter seem to give problematic results \cite{introCDM}, this answer is not definitive yet.\\
A second problem arises from the cosmological constant: if it has always been constant, it means that its density was very small in the early Universe in comparison to the density of the other fluids, and appears to be dominant today only ``by chance''. This is the coincidence problem. To solve this question of coincidence, one prefers to consider that this energy is not constant, and the fluid which was called ``cosmological constant'' is now referred as ``dark energy''. The nature of this dark energy is also unknown and even stranger than usual fluids, as its pressure is negative today. Usual models for such a fluid are the quintessence models, which consider that the behavior of the fluid can be explained with a real scalar field associated to a potential dominating today \cite{introQuint}. However, this potential is unknown, and recent observations of the supernov\ae~of type Ia seem to indicate that the ratio of pressure over density is near to -1, corresponding to a real cosmological constant, and can even be less than -1, what then rules out the involvement of usual real scalar fields and calls for even more exotic explanations.\\
To summarize, the present standard model of cosmology  assumes the existence of two components of completely unknown nature, and the general approach to this assumption is to find constraints to better understand their behaviors. The usual method to find the best values of the different parameters of the model is to try to predict observations and to adjust the parameters to improve the accuracy of the predictions. In this paper, I will consider a model in which the dark matter and the dark energy are in fact different aspects of a same fluid, that I will call ``dark fluid''. Such an idea is simpler because one replaces two components of unknown nature by only one. In the following I will consider constraints from the observations of supernov\ae~of type Ia, cosmic microwave background (CMB), large scale structures, and the theoretical predictions from big-bang nucleosynthesis (BBN), and I will show then that they cannot distinguish between a model based on two components and a model using only one. I will also use the observations to set constraints on the dark fluid. In a last paragraph, I will provide some ideas on what could be the nature of the dark fluid.
%%%%%%%%%%%%%%%%%%%%%%%%%%%%%%%%%%%%%%%%%%%%%%%%%%%%%%%%%%%%%%%%%%%%%%%%%%%%%%%%%%%%%%%%%%%%%%%%%%%%%%%%%%%%%%%%%%%%%%%%
%%%%%%%%%%%%%%%%%%%%%%%%%%%%%%%%%%%%%%%%%%%%%%%%%%%%%%%%%%%%%%%%%%%%%%%%%%%%%%%%%%%%%%%%%%%%%%%%%%%%%%%%%%%%%%%%%%%%%%%%
\section{Basis of the model and definitions}
\noindent If one considers a Friedmann--Lema{\^\i}tre Universe with different fluids: photons, neutrinos, baryons and a dark fluid, the Friedmann equations take the form:
\begin{eqnarray}
\left\{ \frac{\dot{a}}{a} \right\}^2 = H^2 = \frac{8\pi G}{3} \rho - \frac{k}{a^2} \hspace{1.5cm},\hspace{1.5cm} \frac{\ddot{a}}{a} = -\frac{4\pi G}{3} \left\{ \rho + 3 P \right\}\;\; ,
\end{eqnarray}
where $P$ and $\rho$ denote the total pressure and the total density in the Universe, respectively. It will be assumed in the whole article that the value of $a$ today is 1. For the dark fluid model, pressure and density can be expanded as:
\begin{eqnarray}
\nonumber P &=& P_r + P_D \;\; ,\\
\rho &=& \rho_r + \rho_b + \rho_D \;\; ,
\end{eqnarray}
where $r$ denotes the radiation ({\it i.e.} photons + neutrinos), $b$ the baryonic matter and $D$ the dark fluid.\\
In the standard model of cosmology, the equivalent expressions are:
\begin{eqnarray}
\nonumber P &=& P_r + P_\varphi \;\; ,\\
\rho &=& \rho_r + \rho_b + \rho_{dm} + \rho_\varphi \;\; ,
\end{eqnarray}
where $dm$ denotes the dark matter and $\varphi$ the dark energy.\\
For each fluid, one can write the conservation of the energy--momentum tensor in a homogeneous and isotropic spacetime, which reads:
\begin{equation}
\frac{d}{dt}(\rho_{\mbox{fluid}} \, a^3) = - P_{\mbox{fluid}} \, \frac{d(a^3)}{dt} \;\; .
\end{equation}
This equation is equivalent to the second Friedmann equation. As usual, one can define a cosmological parameter corresponding to each fluid by:
\begin{equation}
\Omega^0_{\mbox{fluid}} = \frac{\rho^0_{\mbox{fluid}}}{\rho_0^c} \;\;,
\end{equation}
where $\rho_0^c$ is the critical density today. A cosmological parameter containing the curvature term can also be written:
\begin{equation}
\Omega_K \equiv -\frac{k}{a^2 \; 3 H^2/(8\pi G)} \;\;,
\end{equation}
so that one gets finally a simple equation:
\begin{equation}
\Omega_K + \Omega_r + \Omega_b + \Omega_D = 1 \;\;.
\end{equation}
For a flat Universe, one has $\Omega_K = 0$, and the Friedmann equations are remarkably simplified.\\
Finally, one can define the equation of state of the fluid as:
\begin{equation}
\omega_{\mbox{fluid}} \equiv \frac{P_{\mbox{fluid}}}{\rho_{\mbox{fluid}}} \;\;,
\end{equation}
and one knows that for baryonic matter $\omega_b = 0$, for radiation $\omega_r = 1/3$ and for a real cosmological constant $\omega_\phi = -1$. For simplicity reasons, it is considered in the following that this fluid is perfect, {\it i.e.} the entropy variations and the shear stress can be ignored. For other assumptions, it is necessary to specify a dark fluid model.
\\
A basis for the dark fluid model is now defined, and a comparison with the observations of supernov\ae~of type Ia is given in the following.
%%%%%%%%%%%%%%%%%%%%%%%%%%%%%%%%%%%%%%%%%%%%%%%%%%%%%%%%%%%%%%%%%%%%%%%%%%%%%%%%%%%%%%%%%%%%%%%%%%%%%%%%%%%%%%%
%%%%%%%%%%%%%%%%%%%%%%%%%%%%%%%%%%%%%%%%%%%%%%%%%%%%%%%%%%%%%%%%%%%%%%%%%%%%%%%%%%%%%%%%%%%%%%%%%%%%%%%%%%%%%%%
\section{Constraints from the Supernovae of Type Ia}
\noindent Cosmological constraints from the supernov\ae~of type Ia are based on the joint observations of the redshift $z$ and of the apparent luminosity $l$ of a large number of supernov\ae. Supernov\ae~of type Ia are often considered as standard candles, {\it i.e.} the absolute luminosity $L$ is approximately the same for every supernova (it is not completely true and recent studies correct the value of the absolute luminosity to reflect the deviations from the standard candle behavior \cite{SNeIa1}), so that it is possible to determine for each supernova the luminosity distance
\begin{equation}
d_L=\left(\frac{L}{4\pi l}\right)^{1/2} \;\;.
\end{equation}
This luminosity distance depends on the reddening induced by the expansion of the Universe, and thus can reveal the presence of the cosmological components, through the equation:
\begin{equation}
d_L (z) = \frac{c}{H_0} \frac{1+z}{\sqrt{\left|\Omega_K^0\right|}} S \left\{ \sqrt{\left|\Omega_K^0\right|}
\int^z_0 \frac{dz'}{\sqrt{F(z')}}\right\} \,\, ,
\end{equation}
where $z = a^{-1} - 1$ denotes the redshift, $S$ is given by
\begin{equation}
S ( x ) \equiv \left\{
\begin{array}{ll}
\sin x & \mbox{if $k > 0$} \,\, ,\\
x & \mbox{if $k = 0$} \,\, ,\\
\sinh x & \mbox{if $k < 0$} \,\, ,
\end{array}
\right.
\end{equation}
and $F$ is defined as
\begin{equation}
F(z) = - H_0^{-1} (1+z)^{-2} \,\frac{dz}{dt} \;\;.
\end{equation}
One can see that $F$ is directly related to the first Friedmann equation through the term $dz/dt=-a^{-2} da/dt$.
If there is only one component -- replacing the two dark components -- it shall have the same influence on the luminosity distance -- and then on the expansion of the Universe -- as dark matter and dark energy would have. Through the Friedmann equations, it seems clear that if
\begin{eqnarray}
\nonumber \rho_D&=&\rho_{dm}+\rho_\varphi \;\; ,\\
P_D&=&P_{dm}+P_\varphi=P_\varphi \;\; ,
\end{eqnarray}
the dark fluid would provide the same effect on the expansion of the Universe as the two components.\\
Observations on the supernov\ae~of type Ia enable to give constraints on the dark component densities and on the dark energy behavior at low redshift \cite{SNeIa2}. From these constraints, it should be possible to characterize the dark fluid at low redshift. At first, the cosmological parameter corresponding to the dark fluid can be written in function of those related to dark matter and to dark energy:
\begin{equation}
\Omega_D^0 = \Omega_{dm}^0 + \Omega_\varphi^0 \;\; .
\end{equation}
At low redshift, one can consider that the equation of state for the dark energy is, at first order in~$z$:
\begin{equation}
\omega_\varphi = \omega_\varphi^0 + \omega_\varphi^1 z \;\;,
\end{equation}
and that the equation of state for the dark fluid can have the same form:
\begin{equation}
\omega_D = \omega_D^0 + \omega_D^1 z \;\;.
\end{equation}
The observations have given constraints on the values of $\omega_\varphi^0$ and $\omega_\varphi^1$, and one would like to deduce from them constraints on $\omega_D^0$ and $\omega_D^1$. The equation of state of the dark fluid writes:
\begin{equation}
\omega_D=\frac{P_D}{\rho_D}=\frac{P_\varphi}{\rho_{dm}+\rho_\varphi}=\omega_\varphi \frac{\rho_\varphi}{\rho_{dm}+\rho_\varphi} \;\;.
\end{equation}
At first order in $z$, the dark matter density evolves like $\rho_{dm}=\rho^0_{dm} a^{-3} = \rho^0_{dm} (1+3z)$ . It would be interesting to know the behavior of $\rho_\varphi$. Let us assume that, at first order:
\begin{equation}
\rho_\varphi=\rho_\varphi^0+\rho_\varphi^1 z \;\;.
\end{equation}
The equation of conservation of the energy-momentum tensor for each fluid satisfies:
\begin{equation}
\frac{d}{dt}(\rho_\varphi a^3) = -P_\varphi \frac{d(a^3)}{dt} \;\;.
\end{equation}
At first order in $z$, this equation becomes:
\begin{equation}
\frac{d}{dt}(\rho_\varphi^0 + z (\rho_\varphi^1 - 3\rho_\varphi^0)) = - (\omega_\varphi^0
\rho_\varphi^0) \frac{d(1-3z)}{dt} \;\;,
\end{equation}
so that the density of dark energy reads:
\begin{equation}
\rho_\varphi^1 = 3 \rho_\varphi^0 (1+\omega_\varphi^0) \;\;.
\end{equation}
Then, the relation between the ratio pressure/density for the dark fluid becomes:
\begin{equation}
\omega_D=(\omega_\varphi^0+\omega_\varphi^1 z) \frac{\rho_\varphi^0 (1 + 3 (1+\omega_\varphi^0) z)}{\rho_{dm}^0 (1+3z) +\rho_\varphi^0 (1 + 3 (1+\omega_\varphi^0) z)}\;\;,
\end{equation}
and one can determine the value of the two first terms of the expansion:
\begin{eqnarray}
\nonumber \omega_D^0&=& \frac{\omega_\varphi^0 \Omega^0_\varphi}{\Omega^0_{dm} + \Omega^0_\varphi}\;\;,\\
\omega_D^1&=& \frac{\omega_\varphi^1 \Omega^0_\varphi}{\Omega^0_{dm} + \Omega^0_\varphi} + \frac{3 \Omega^0_{dm}\Omega^0_\varphi (\omega_\varphi^0)^2}{(\Omega^0_{dm} + \Omega^0_\varphi)^2}\;\;.
\end{eqnarray}
The favored values for the cosmological parameters of the usual standard model from the supernov\ae~of type Ia \cite{SNeIa2} combined with the results of other observations \cite{wmap1} are:
\begin{eqnarray}
\nonumber h =& 0.70 \pm 0.04\\
\Omega_K^0 = -0.012 \pm 0.022 & & \Omega_{dm}^0 = 0.25 \pm 0.04\\
\nonumber \Omega_b^0 = 0.049 \pm 0.012  & & \nonumber \Omega_\varphi^0 = 0.712 \pm 0.044\\
\nonumber \omega_\varphi^0 = -1.02 \pm 0.19 & & \omega_\varphi^1 = 0.6 \pm 0.5 \;\;.
\end{eqnarray}
From these values, one can calculate the parameters of the dark fluid:
\begin{eqnarray}
\nonumber \Omega_D^0 &=& 0.962 \pm 0.084\\
\label{resultSN}
\omega_D^0 &=& -0.76 \pm 0.25\\
\nonumber \omega_D^1 &=&  1.0 \pm 0.6 \;\;.
\end{eqnarray}
These values are of course not completely representative of the dark fluid model, because they come from data analyses based on the usual standard model. Nevertheless, one can use them as test--parameters at low redshift.\\
Recent supernova observations tend to show that the dark energy has a negative pressure. Moreover, $\omega_\varphi<-1$ is not at all excluded, and in that case the dark energy cannot be explained anymore thanks to the usual models (see for example \cite{caldwell} for a possible answer to this problem). From the precedent constraints, one can see that this difficulty vanishes with a dark fluid. Hence, the pressure of the fluid has to be negative today at cosmological scales, and seems to increase strongly with the redshift. As $\omega_D^0 \ge -1$, it seems possible to model the dark fluid with a scalar field.\\
The study of supernov\ae~provided us properties of the equation of state of our dark fluid at low redshift independently from the specification of a dark fluid model. We will now try to extract constraints from the information concerning large scale structures.
%%%%%%%%%%%%%%%%%%%%%%%%%%%%%%%%%%%%%%%%%%%%%%%%%%%%%%%%%%%%%%%%%%%%%%%%%%%%%%%%%%%%%%%%%%%%%%%%%%%%%%%%%%%%%%%
%%%%%%%%%%%%%%%%%%%%%%%%%%%%%%%%%%%%%%%%%%%%%%%%%%%%%%%%%%%%%%%%%%%%%%%%%%%%%%%%%%%%%%%%%%%%%%%%%%%%%%%%%%%%%%%
\section{Large Scale Structures}
\noindent We will not consider here a complete scenario of structure formation, which would require the specification of a precise model of dark fluid. Nevertheless, one can study the necessary conditions for the fluid parameters to enable the perturbations to grow and to give birth to large scale structures.\\
Let us consider the case where the equation of state of our fluid does not change during the growth of perturbations, and, to simplify, that the entropy perturbations can be ignored and that the Jeans length is smaller than any other considered scale. One can define the local density contrast of the dark fluid as:
\begin{equation}
\delta\left(\vec{x},t\right) \equiv \frac{\rho_D\left(\vec{x},t)\right)}{\overline{\rho_D}(t)} - 1 \;\;,
\end{equation}
where $\rho_D\left(\vec{x},t\right)$ is the local value of the density, and $\overline{\rho_D}(t)$ is the mean background density, {\it i.e.} the apparent cosmological density. In the fluid approximation, one can write the evolution equation of the local density contrast \cite{LSS}:
\begin{eqnarray}
\label{eq.evol}
\nonumber &\displaystyle \frac{1}{H^2}\frac{d^2\delta}{dt^2} + \left(2+ \frac{\dot{H}}{H^2}\right) \frac{1}{H}\frac{d\delta}{dt} - \frac{2}{3} (1+\omega_D)(1+3\omega_D) \Omega_D \delta\\
&\displaystyle = \frac{4+3\omega_D}{3(1+\omega_D)} \frac{1}{1+\delta} \left( \frac{1}{1+\delta}\right)^2 \frac{1}{H^2} \left( \frac{d\delta}{dt} \right)^2 + \frac{3}{2} (1+\omega_D) (1+3\omega_D) \Omega_D \delta^2\;\; .
\end{eqnarray}
To solve this equation, one can define a new variable reflecting the expansion:
\begin{equation}
\eta = \ln a \;\;,
\end{equation}
so that, if one assumes that the dark fluid is completely dominant at the time of growth of perturbations (in that case, the Friedmann equations reveal that $\dot{H}/H^2=-3(1+\omega_D)/2$), equation (\ref{eq.evol}) becomes:
\begin{equation}
\label{eq.fluid}
\frac{d^2\delta}{d\eta^2} + \frac{1- 3 \omega_D}{2} \frac{d\delta}{d\eta} - \frac{2}{3} (1+\omega_D)(1+3\omega_D) \delta = \frac{4+3\omega_D}{3(1+\omega_D)} \frac{1}{1+\delta} \left( \frac{1}{1+\delta}\right)^2 \left( \frac{d\delta}{d\eta} \right)^2 + \frac{3}{2} (1+\omega_D) (1+3\omega_D) \delta^2\;\; .
\end{equation}
Because the coefficients of the above equation are time--dependant only, one can separate the spatial and temporal parts so that
\begin{equation}
\delta_l\left(\vec{x},t\right)=\delta_0\left(\vec{x}\right) D(t)\;\;,
\end{equation}
where $D$ is called the ``linear growth factor''. In the linear approximation, where $\delta$ is small, equation (\ref{eq.fluid}) becomes:
\begin{equation}
\frac{d^2 D}{d\eta^2} + \frac{1 - 3 \omega_D}{2} \frac{dD}{d\eta} - \frac{2}{3} (1+\omega_D)(1+3\omega_D) D = 0 \;\;.
\end{equation}
Its solutions take the form $D= D_1 a^{\alpha_1} + D_2 a^{\alpha_2}$, with $D_1$ and $D_2$ being two integration constants, and
\begin{eqnarray}
\alpha_1 &=& 1 + 3 \omega_D\;\;,\\
\alpha_2 &=& - \frac{3}{2} (1+\omega_D) \;\;.
\end{eqnarray}
Thus, in the case of a dominant dark fluid, we have only a growing mode if $\omega_D > -1/3$ or if $\omega_D < -1$. One can however note that the last inequality seems very difficult to achieve with standard model for dark matter or dark energy models.\\
Let us now consider the observations of the cosmic microwave background to get constraints at earlier times.
%%%%%%%%%%%%%%%%%%%%%%%%%%%%%%%%%%%%%%%%%%%%%%%%%%%%%%%%%%%%%%%%%%%%%%%%%%%%%%%%%%%%%%%%%%%%%%%%%%%%%%%%%%%%%%%%%%
%%%%%%%%%%%%%%%%%%%%%%%%%%%%%%%%%%%%%%%%%%%%%%%%%%%%%%%%%%%%%%%%%%%%%%%%%%%%%%%%%%%%%%%%%%%%%%%%%%%%%%%%%%%%%%%%%%
\section{Cosmic Microwave Background}
\noindent A power spectrum of temperature fluctuations can be deduced from the observations of the cosmic microwave background (CMB) \cite{wmap2}. Predicting this power spectrum requires a hard work, and a program like CMBFAST \cite{cmbfast} is able to produce it for the cosmological standard model. In our case, we will consider only the position of the peaks to constrain the parameters of the dark fluid, and we will make some assumptions on the dark fluid properties.\\
First, one should note that at high redshift, in the standard model the density of dark energy is nearly negligible in comparison to that of the dark matter. As the usual model seems to be able to correctly reproduce the fluctuations of the CMB, one can assume that our dark fluid should not behave very differently from the superposition dark matter/dark energy, and so should behave at the moment of recombination nearly like matter. Therefore, one can write the density of our fluid as a sum of a matter--like term ($m$) and of another term of unknown behavior ($o$):
\begin{equation}
\rho_D = \rho_{Dm}^{ls} \left(\frac{a}{a_{ls}}\right)^{-3} +  \rho_{Do} \;\;.
\end{equation}
One can note that this equation gives no constraint on the behavior of the dark fluid, as the second term is not restricted to any behavior yet.\\
We do not want to specify a model of dark fluid and we would like to be as general as possible. Nevertheless, we will consider for simplicity only the background properties of the dark fluid, and we will not try to reproduce the whole power spectrum, but only consider the position of its peaks without trying to find their amplitude. The conformal time is defined by:
\begin{equation}
\tau = \int dt \,\, a^{-1}(t) \;\;.
\end{equation}
The spacing between the peaks is then given, to a good approximation, by \cite{hu_sugiyama}:
\begin{equation}
\Delta l \approx \pi \frac{\tau_0 - \tau_{ls}}{\overline{c}_s \tau_{ls}} \;\;,
\end{equation}
where $\overline{c}_s$ is the average sound speed before last scattering, and $\tau_0$ and $\tau_{ls}$ the conformal time today and at last scattering. This average sound speed reads:
\begin{equation}
\overline{c}_s \equiv \tau_{ls}^{-1} \int_0^{\tau_{ls}} d\tau \left( 3 + \frac{9 \rho_b(t)}{4 \rho_r(t)}\right)^{-1/2} \;\;,
\end{equation}
where $\rho_b$ is the density of baryonic matter and $\rho_r$ is the density of relativistic fluids (radiation and neutrinos).\\
\\
Let us consider that the Universe is flat so that the Friedmann equations are simplified. In this case, the first Friedmann equation can be written:
\begin{equation}
H^2=\frac{8\pi G}{3} \left( \rho_b + \rho_r + \rho_{Dm} + \rho_{Do} \right) \;\;.
\end{equation}
Using the evolution equation of the different densities, this becomes:
\begin{equation}
H^2= H_0^2 \left( \Omega^0_b a^{-3} + \Omega^0_r a^{-4} + \Omega^{ls}_{Dm} \left(\frac{a}{a_{ls}}\right)^{-3} \right) + \frac{8\pi G}{3} \rho_{Do} \;\;.
\end{equation}
The precedent equation cannot be solved if the form of $\rho_{Do}$ is not given. In our case, it is possible to assume that the fraction
\begin{equation}
\Omega_{Do} (\tau) \equiv \frac{\rho_{Do}(\tau)}{\sum \rho(\tau)}
\end{equation}
does not vary too rapidly before the moment of last scattering (denoted $ls$), so that an effective average can be defined:
\begin{equation}
\overline{\Omega}^{ls}_{Do} \equiv \tau_{ls}^{-1} \int_0^{\tau_{ls}} \Omega_{Do} (\tau) d\tau \;\;.
\end{equation}
In the following, we will only consider an approximate and effective density:
\begin{equation}
\rho_{Do} \approx H^2 \frac{3}{8\pi G} \overline{\Omega}^{ls}_{Do} \;\;.
\end{equation}
Let us then replace this density in the Friedmann equation:
\begin{equation}
H^2 (1 - \overline{\Omega}^{ls}_{Do}) = H_0^2 \left( ( \Omega^0_b + \Omega^{ls}_{Dm} a_{ls}^3) a^{-3} + \Omega^0_r a^{-4} \right) \;\;.
\end{equation}
This time, provided one fixes the values of the different cosmological parameters and knowing the initial conditions, this equation can be solved. While replacing usual time by conformal time, the Friedmann equation becomes:
\begin{equation}
\left( \frac{da}{d\tau} \right)^2  = H_0^2 (1 - \overline{\Omega}^{ls}_{Do})^{-1} \left( ( \Omega^0_b + \Omega^{ls}_{Dm} a_{ls}^3) a(\tau)+ \Omega^0_r \right) \;\;.
\end{equation}
The value of the conformal time at the moment of last scattering is given by:
\begin{equation}
\tau_{ls} = 2 H_0^{-1} \sqrt{\frac{1 - \overline{\Omega}^{ls}_{Do}}{\Omega^0_b + \Omega^{ls}_{Dm} a_{ls}^3}} \left\{ \sqrt{a_{ls} + \frac{\Omega^0_r}{\Omega^0_b +
\Omega^{ls}_{Dm} a_{ls}^3}} - \sqrt{\frac{\Omega^0_r}{\Omega^0_b +
\Omega^{ls}_{Dm} a_{ls}^3}}\right\} \;\;.
\end{equation}
One can use the same method to evaluate the conformal time today. The Friedmann equation reads, after last scattering:
\begin{equation}
\left( \frac{da}{d\tau} \right)^2 = H_0^2 \left( \Omega^0_b a(\tau) + \Omega^0_r + a(\tau)^4 \;\; \frac{\rho_{D}}{\rho_0^C}\right)\;\;.
\end{equation}
Let us make now the further assumption that the dark fluid has an approximate equation of state:
\begin{equation}
\rho_D = \tilde{\rho}_D^0 a^{-3(1+\overline{\omega}_D)} \;\;,
\end{equation}
where $\tilde{\rho}_D^0$ is an effective value of the dark fluid density, such that
\begin{equation}
\tilde{\rho}_D^0 = a_{ls}^{3(1+\overline{\omega}_D)} \rho_D^{ls} \;\;,
\end{equation}
and $\overline{\omega}_D$ is the average value of $\omega_D$ over the conformal time, weighted by:
\begin{equation}
\Omega_D(\tau) = \frac{\rho_D(\tau)}{\sum \rho(\tau)}
\end{equation}
to reflect the fact that the equation of state of our fluid should be more significant when its density contributes more heavily to the total density of the Universe. $\overline{\omega}_D$ is then given by:
\begin{equation}
\overline{\omega}_D \equiv \frac{\displaystyle \int_0^{\tau_0} \Omega_D(\tau) \omega_D(\tau) d\tau}{\displaystyle \int_0^{\tau_0} \Omega_D(\tau) d\tau} \;\;.
\end{equation}
Defining the effective cosmological parameter $\tilde{\Omega}^0_D=3 \tilde{\rho}^0_D / (8\pi G H_0^2)$, the Friedmann equation becomes:
\begin{equation}
\label{eq.omegatilde}
\left( \frac{da}{d\tau} \right)^2 = H_0^2 \left( \Omega^0_b a(\tau) + \Omega^0_r + \tilde{\Omega}^0_D a(\tau)^{(1-3\overline{\omega}_D)} \right) \;\;.
\end{equation}
One can then integrate the equation, and show that:
\begin{equation}
\tau_{0} = 2 H_0^{-1} F(\overline{\omega}_D) \;\;,
\end{equation}
with
\begin{equation}
F(\overline{\omega}_D) \equiv \frac{1}{2} \int_0^1 da \left( \Omega^0_b a + \Omega^0_r + \tilde{\Omega}^0_D a^{(1-3\overline{\omega}_D)} \right)^{-1/2} \;\;.
\end{equation}
There is no analytical integration for this function, but in a few cases. In particular, one has for a flat Universe:
\begin{equation}
F(0) = \frac{1}{\Omega_b^0+\tilde{\Omega}^0_D}\left( \sqrt{\Omega^0_b + \Omega^0_r + \tilde{\Omega}^0_D} - \sqrt{\Omega_r^0} \right) \;\;,
\end{equation}
and
\begin{equation}
F(-1/3) = \frac{1}{2} (\Omega^0_D)^{-1/2} \ln \left( \frac{1-\tilde{\Omega}^0_r + 2 \sqrt{\Omega^0_D}}{\Omega^0_b + 2 \sqrt{\Omega^0_r\Omega^0_D}}\right) \;\;.
\end{equation}
The case $\overline{\omega}_D=-1/3$ looks much more probable than $\overline{\omega}_D=0$ when one considers the value of $\omega^0_D$ which was obtained from the supernova data. One finally gets the spacing between peaks:
\begin{equation}
\Delta l=\pi \overline{c}_s^{-1} \left[ F(\overline{\omega}_D) \sqrt{\frac{\Omega^0_b + \Omega^{ls}_{Dm} a_{ls}^3}{1 - \overline{\Omega}^{ls}_{Do}}} \left\{ \sqrt{a_{ls} + \frac{\Omega^0_r}{\Omega^0_b +
\Omega^{ls}_{Dm} a_{ls}^3}} - \sqrt{\frac{\Omega^0_r}{\Omega^0_b +
\Omega^{ls}_{Dm} a_{ls}^3}}\right\}^{-1} -1\right] \;\;.
\end{equation}
The sound velocity $\overline{c}_s$ is then given by:
\begin{equation}
\overline{c}_s = \tau_{ls}^{-1} H_0^{-1} \sqrt{1 - \overline{\Omega}^{ls}_{Do}} \int_0^{a_{ls}} da \left[ \left(3 + \frac{9 \Omega_b^0}{4 \Omega_r^0} a \right)  \left( ( \Omega^0_b + \Omega^{ls}_{Dm} a_{ls}^3) a+ \Omega^0_r \right)\right]^{-1/2} \;\;.
\end{equation}
This equation can be integrated analytically, and one finally gets:
\begin{eqnarray}
\overline{c}_s &=& \frac{1}{3} \left(\frac{\Omega_r^0}{\Omega^0_b}\right)^{1/2} \left\{ \sqrt{a_{ls} + \frac{\Omega^0_r}{\Omega^0_b +
\Omega^{ls}_{Dm} a_{ls}^3}} - \sqrt{\frac{\Omega^0_r}{\Omega^0_b +\Omega^{ls}_{Dm} a_{ls}^3}} \right\}^{-1} \times\\
\nonumber&&\hspace{-2.5cm}\ln\left(\frac{\Omega_r^0 (7\Omega_b^0+4\Omega^{ls}_{Dm} a_{ls}^3) + 2\sqrt{3\Omega_b^0 (\Omega^0_b + \Omega^{ls}_{Dm} a_{ls}^3)(\Omega^0_b a_{ls}+
\Omega^{ls}_{Dm} a_{ls}^4 + \Omega^0_r)(3 \Omega^0_b a_{ls} + 4\Omega^0_r)}+6 a_{ls}\Omega^0_b(\Omega^0_b + \Omega^{ls}_{Dm} a_{ls}^3)}{\Omega_r^0 (7\omega_b^0+4\Omega^{ls}_{Dm} a_{ls}^3)+ 4 \Omega_r^0 \sqrt{3\Omega_b^0 (\Omega^0_b +
\Omega^{ls}_{Dm} a_{ls}^3)}} \right) \;\;.
\end{eqnarray}
One can notice that $\overline{c}_s$ does not depend on $\overline{\Omega}^{ls}_{Do}$. The approximate value of $a_{ls}$ can be taken from~\cite{hu_fukugita}:
\begin{equation}
a_{ls}^{-1} \approx 1008 (1+0.00124 (\Omega_b^0 h^2)^{-0.74}) (1+c_1 (\Omega^{ls}_{Dm} a_{ls}^3)^{c_2}) \;\;,
\end{equation}
where
\begin{eqnarray}
\nonumber c_1 &=& 0.0783 (\Omega_b^0 h^2)^{-0.24}(1+39.5 (\Omega_b^0 h^2)^{0.76})^{-1}\;\;,\\
c_2 &=& 0.56 (1+21.1 (\Omega_b^0 h^2)^{1.28})^{-1} \;\;.
\end{eqnarray}
Unfortunately, this is an implicit equation, and we cannot find $a_{ls}$ analytically.\\
Finally, one has a direct dependence between $\Delta l$ and the parameters $\Omega^{ls}_{Dm}$, $\overline{\Omega}^{ls}_{Do}$, $\tilde{\Omega}_D^0$ and $\overline{\omega}_D$. Even if this is only an approximate formula, it gives the possibility to visualize directly the effect of the dark fluid on the position of the peaks, provided that its behavior does not differ much from the requirements of the approximations on $\overline{\Omega}^{ls}_{Do}$ and $\overline{\omega}_D$.\\
However, the calculated $\Delta l$ cannot be directly related to the observed spacing between peaks, as shifts of peaks can be induced by other effects. In particular, the location of the i--th peaks can be approximated by:
\begin{equation}
l_i=\Delta l (m-\phi_i)=\Delta l (m-\bar\phi-\delta\phi_i) \;\;,
\end{equation}
where $\bar\phi$ is the shift of the first peak, corresponding to an overall shift, and the $\delta\phi_i$ is the specific shift of the i--th peak. We have fortunately access to the fitting formulae of \cite{doran_lilley}:\\
\\
\underline{Overall phase shift $\bar\phi$ :}
\begin{equation}
\bar{\phi} =(1.466 - 0.466 n) \left[ a_1 r_{ls}^{a_2} + 0.291 \overline{\Omega}^{ls}_{Do} \right] \;\;,
\end{equation}
where $a_1$ and $a_2$ are given by:
\begin{eqnarray}
\nonumber a_1 & = & 0.286 + 0.626\left(\Omega_b h^2\right) \;\;,\\
a_2 & = & 0.1786 - 6.308 \Omega_b h^2 + 174.9\left(\Omega_b h^2\right)^2 - 1168 \left(\Omega_b h^2\right)^3 \;\;,
\end{eqnarray}
$n$ is the spectral index and $r_{ls}$ is defined by:
\begin{equation}
r_{ls} = \frac{\rho_r(a_{ls})}{\rho_{Dm}(a_{ls})} = \frac{\Omega_r^0}{\Omega_{Dm}^{ls} a_{ls}^4} \;\;.
\end{equation}
\underline{Relative shift of second peak $\delta\phi_2$ :}
\begin{equation}
\delta\phi_2 = c_0 - c_1 r_{ls} - c_2 r_{ls}^{-c_3} + 0.05\,(n-1) \;\;,
\end{equation}
with
\begin{eqnarray}
\nonumber c_0 &=& -0.1 + \left( 0.213 - 0.123 \overline{\Omega}^{ls}_{Do} \right) \exp\left\{ - \left( 52 - 63.6 \overline{\Omega}^{ls}_{Do}\right) \Omega_b h^2 \right\}\;\;,\\
c_1 &=& 0.063 \,\exp\left \{-3500 \left(\Omega_b h^2\right)^2\right\} +0.015\;\;,\\
\nonumber c_2 &=& 6\times 10^{-6} + 0.137 \left (\Omega_b h^2 - 0.07 \right) ^2\;\;,\\
\nonumber c_3 &=& 0.8 + 2.3 \overline{\Omega}^{ls}_{Do} + \left( 70 - 126 \overline{\Omega}^{ls}_{Do}\right) \Omega_b h^2 \;\;.
\end{eqnarray}
\underline{Relative shift of third peak $\delta\phi_3$ :}
\begin{equation}
\delta\phi_3 = 10 - d_1 r_{ls}^{d_2} + 0.08\, (n-1) \;\;,
\end{equation}
with
\begin{eqnarray}
d_1 &=& 9.97 + \left(3.3 -3 \overline{\Omega}^{ls}_{Do}\right) \Omega_b h^2 \;\;,\\
\nonumber d_2 &=& 0.0016 - 0.0067 \overline{\Omega}^{ls}_{Do}  + \left(0.196 - 0.22 \overline{\Omega}^{ls}_{Do}\right) \Omega_b h^2 + \frac{(2.25 + 2.77  \overline{\Omega}^{ls}_{Do} ) \times 10^{-5}}{\Omega_b h^2} \;\;.
\end{eqnarray}
One can now compare our results to the data. The WMAP experiment provides the precise location of the two first peaks \cite{wmap2}:
\begin{eqnarray}
\nonumber l_{p_1} &=& 220.1 \pm 0.8\;\;,\\
l_{p_2} &=& 546 \pm 10 \;\;,
\end{eqnarray}
and BOOMERanG gives the position of the third peak \cite{boomerang}:
\begin{equation}
l_{p_3} = 825 \pm 13 \;\;.
\end{equation}
To evaluate roughly the value of the parameters of the fluid, one can fix the other parameters as follows:
\begin{eqnarray}
\nonumber n &=& 1\;\;,\\
h &=& 0.70\;\;,\\
\nonumber \Omega_b^0 &=& 0.049\;\;,\\
\nonumber\Omega_r^0 &=& 9.89 \times 10^{-5} \;\;.
\end{eqnarray}
For these values, one finds in Table 1 the resulting positions of the peaks in function of parameters of the dark fluid.
\begin{table*}
\begin{center}
\begin{tabular}{ccccccc} \hline\hline
$\Omega^{ls}_{Dm}$ & $\overline{\Omega}^{ls}_{Do}$ & $\tilde{\Omega}_D^0$ & $\overline{\omega}_D$ & $l_1$ & $l_2$ & $l_3$\\
\hline
0.600~~ & ~~0.125~~ & ~~0.95~~ & ~~-0.10~~~ & ~~~220.3~~ & ~~544.8~~ & ~~817.2\\
0.400~~ & ~~0.120~~ & ~~0.95~~ & ~~-0.15~~~ & ~~~220.0~~ & ~~546.1~~ & ~~830.6\\
0.270~~ & ~~0.080~~ & ~~0.95~~ & ~~-0.20~~~ & ~~~219.4~~ & ~~544.8~~ & ~~835.4\\
0.200~~ & ~~0.010~~ & ~~0.50~~ & ~~-0.13~~~ & ~~~220.2~~ & ~~544.5~~ & ~~835.9\\
0.250~~ & ~~0.035~~ & ~~0.50~~ & ~~-0.10~~~ & ~~~220.7~~ & ~~544.1~~ & ~~833.5\\
0.018~~ & ~~0.013~~ & ~~0.25~~ & ~~-0.07~~~ & ~~~219.6~~ & ~~543.4~~ & ~~837.4\\
0.017~~ & ~~0.001~~ & ~~0.10~~ & ~~~0.17~~~ & ~~~219.5~~ & ~~543.1~~ & ~~837.6\\
\hline
\end{tabular}
\caption{Position of the peaks for $h =0.7$, $n=1$, $\Omega_b^0 = 0.049$ and $\Omega_r^0=9.89 \times 10^{-5}$, in function of the parameters of the dark fluid.}
\end{center}
\end{table*}\\
\noindent One can note that a large range of values is possible. It is not so strange because we have many parameters for our fluid. A more complete analysis is not needed here, because the strongest constraints would come from the specification of a model. Without specifying a model, Table 1 shows that the values of the parameters of the dark fluid are not stringently constrained. One can nevertheless see that for large values of $\tilde{\Omega}_D^0$, the permitted values of $\overline{\omega}_D$ are negative, and hence in that case one can assume that our fluid behaves today like a cosmological constant whereas it could have behaved mainly like matter at last scattering. For small values of $\tilde{\Omega}_D^0$, $\overline{\omega}_D$ is positive, so that the density of dark fluid should decrease more rapidly than a matter density after last scattering. In this case, $\overline{\Omega}^{ls}_{Do}$ is very small, and the fluid should have behaved like matter before and around last scattering. Small values of $\tilde{\Omega}_D^0$ look therefore unrealistic, because as $\overline{\omega}_D$ is then positive, unless our dark fluid has an oscillating density, $\overline{\Omega}^{ls}_{Do}$ should be much larger and certainly dominant.\\
The value of $\overline{\Omega}^{ls}_{Do}$, which can be as much as 0.1, also shows that before recombination, the fluid may have behaved differently from matter, and perhaps like radiation. One can also note that low values of $\tilde{\Omega}_D^0$ corresponds to low values of $\overline{\Omega}^{ls}_{Do}$, and then in that case the fluid mostly behaves like matter.\\
\\
If one combines these results with the results from the CMB, it seems that the constraints become $-1/3 < \omega_D < 0$ to enable the perturbations to grow. This shows in fact that during the growth of the perturbations the behavior of the dark fluid should not be too different from that of matter. This result also confirms that the value of the effective $\tilde{\Omega}_D^0$ which appears in equation (\ref{eq.omegatilde}) cannot be too small (it has to be at least larger than 0.2). \\
If one wants to perform a much more precise study of a specified model, it would be interesting to simulate the whole process of structure formation, and to compare the results with surveys like SDSS \cite{SDSS}, or 2dF \cite{2dF}. We will not study here further the CMB power spectrum, as other features seem more model--dependant, and we want to consider here the general case. We will now consider the results of the big-bang nucleosynthesis and their influence on the establishment of a dark fluid model.

%%%%%%%%%%%%%%%%%%%%%%%%%%%%%%%%%%%%%%%%%%%%%%%%%%%%%%%%%%%%%%%%%%%%%%%%%%%%%%%%%%%%%%%%%%%%%%%%%%%%%%%%%%%%%%%%%%%%
%%%%%%%%%%%%%%%%%%%%%%%%%%%%%%%%%%%%%%%%%%%%%%%%%%%%%%%%%%%%%%%%%%%%%%%%%%%%%%%%%%%%%%%%%%%%%%%%%%%%%%%%%%%%%%%%%%%%

\section{Big-Bang Nucleosynthesis}
\noindent Recent analyses of the big--bang nucleosynthesis (BBN) \cite{BBN1} indicate a discrepancy between the value of the baryonic density calculated from the observed Li and $^4$He abundances, and the one calculated with the observations of deuterium. Some explanations can be found. Problems could have appeared in the measurement of Li and $^4$He abundances, or the Li on the stellar surface could be altered during stellar evolution, or we have no accurate knowledge of the reaction rates related to $^7$Be destruction, or the expansion rate during BBN could have been modified through an accelerated cosmological expansion. Thus, two possibilities can be considered for the equation of state of the dark fluid. First, if it is correct to consider a Universe dominated by radiation at BBN time, the main constraint is that the dark fluid density should be small in comparison to the radiation density; otherwise Friedmann equations indicates that the expansion rate of the Universe would be different from the one in the standard BBN, changing then the temperature evolution rate and so the abundance of the elements. It means that, if one assumes that the dark fluid behavior does not change violently during BBN, the equation of state of this fluid around the time of BBN has to be $\omega_D(\mbox{BBN}) \leq 1/3$, or that its density was completely negligible before BBN. In the case of a real radiative behavior $\omega_D(\mbox{BBN}) = 1/3$, the dark fluid behaves like extra--families of neutrinos, and its density can be constrained. The effective extra--neutrinos number at the BBN time is defined by:
\begin{equation}
\Delta N_{\mbox{eff}}(\mbox{BBN}) \equiv \frac{\rho_D(\mbox{BBN})}{\rho_\nu(\mbox{BBN})} \;\;,
\end{equation}
where $\rho_\nu$ is the standard density of a single relativistic neutrino species. The usual bound on the number of neutrinos is $\Delta N_{\mbox{eff}} < 1$ \cite{BBN2}, which corresponds in our case for a temperature around 1~MeV ($a^{-1} \approx 3-4 \times 10^9$) to:
\begin{equation}
\rho_D(\mbox{BBN}) < \frac{7}{8} \left( \frac{4}{11}\right)^{4/3} \frac{\pi^2}{15} T^4_{\mbox{BBN}} \approx 3 \times 10^{-2} (\mbox{MeV})^4 \;\;.
\end{equation}
This limit is only valid in the case $\omega_D(\mbox{BBN}) = 1/3$.\\
If the abundance of the elements is as observed, a modification of the expansion rate could provide, as presented in \cite{salati}, a correction to the predicted values. If our fluid is the dominant component at the time of the BBN, it can have a big influence on the expansion rate of the Universe. Evaluating the density of the dark fluid and its evolution during the time of the BBN so that the observations are retrieved would require further studies that I will not develop here.\\
\\
We have seen that a dark fluid may be compatible with the cosmological observations, and could be an interesting approach to the ambivalence of dark energy and dark matter. Let us now consider possible paths to model a dark fluid.
%%%%%%%%%%%%%%%%%%%%%%%%%%%%%%%%%%%%%%%%%%%%%%%%%%%%%%%%%%%%%%%%%%%%%%%%%%%%%%%%%%%%%%%%%%%%%%%%%%%%%%%%%%%%%%%
%%%%%%%%%%%%%%%%%%%%%%%%%%%%%%%%%%%%%%%%%%%%%%%%%%%%%%%%%%%%%%%%%%%%%%%%%%%%%%%%%%%%%%%%%%%%%%%%%%%%%%%%%%%%%%%
\section{Models of Dark Fluid}
\noindent In the literature, only few fluids behaving like a dark fluid are considered. Different ways to model the dark fluid are possible, and I will consider here in particular two of them: the generalized Chaplygin gas, based on D-brane theories, and another one using scalar fields.
\subsection{Generalized Chaplygin Gas}
\noindent The generalized Chaplygin gas (GCG) is an exotic fluid derived from D-brane theories \cite{chaplygin1}. It can be described by an equation of state:
\begin{equation}
P_{ch} = - \frac{A}{\rho^\alpha_{ch}} \;\;,
\end{equation}
where $\alpha$ is a constant, $0 < \alpha \le 1$, and $A$ is another positive constant. This equation of state corresponds to a density evolving like:
\begin{equation}
\rho_{ch} = \rho_{ch}^0 \left( A_s + \frac{1-A_s}{a^{3(1+\alpha)}}\right)^{1/(1+\alpha)} \;\;,
\end{equation}
where $A_s=A/(\rho_{ch}^0)^{(1+\alpha)}$ and $\rho_{ch}^0$ is the Chaplygin gas density today. Such a behavior could be interesting in order to model the dark fluid, because for high values of $a$ this density is mainly constant, and for low values of $a$ it evolves like matter. This behavior has to be compared with the observations.\\
For the comparison with the data of supernov\ae~of type Ia, let us derive the equation of state of the GCG at low redshift:
\begin{equation}
\omega_{ch}=\frac{P_{ch}}{\rho_{ch}} = - \frac{A_s}{A_s+(1-A_s) (1+z)^{3(1+\alpha)}} \approx -A_s \left[ 1 - 3 (1-A_s) (1+\alpha)z\right] \;\;,
\end{equation}
so that one can deduce:
\begin{eqnarray}
\nonumber \omega^0_{ch}&=&-A_s \;\;,\\
\omega^1_{ch}&=& 3 A_s (1-A_s) (1+\alpha) \;\;,
\end{eqnarray}
to be compared with the constraints (\ref{resultSN}), and one finally gets:
\begin{eqnarray}
\nonumber As & = & 0.76 \pm 0.25\;\;,\\
\alpha & = & 0.8 \pm 2.4 \;\;.
\end{eqnarray}
We have of course no constraint on $\alpha$. The analysis of CMB has been done in \cite{chaplygin2}, and the results for $h=0.7$ and $n=1$, combined with the ones from the supernov\ae~of type Ia are:
\begin{eqnarray}
As & = & 0.75 \pm 0.13\;\;,\\
\nonumber \alpha & = & 0.4 \pm 0.2 \;\;.
\end{eqnarray}
If one considers now the BBN, at this time the GCG behaves like matter, so that it is compatible with the standard BBN scenario. The large scale structure formation has been studied in a case where the Chaplygin gas adds only a background density to a Universe containing cold dark matter \cite{chaplygin3}, and such a scenario seems then possible.\\
So, the GCG seems to be in agreement with the cosmological observations. A further analysis is still needed, in particular concerning the growth of structures with a dominant Chaplygin gas density, or the local behavior of such a fluid.
\subsection{Scalar Fields}
\noindent One can also consider the idea that the dark fluid could be explained thanks to a scalar field. Indeed, scalar fields are very useful in explaining the behavior of the dark energy today \cite{quintessence1,quintessence2}, and recent analyses have shown that they can behave like matter on local scales \cite{arbey1,sflocal} as well as on cosmological scales \cite{arbey2,sfcosmo}.\\
Let us therefore consider a real scalar field associated with a Lagrangian density
\begin{equation}
\mathcal{L} \; = \; g^{\mu \nu} \, \partial_{\mu} \varphi \, \partial_{\nu} \varphi \; - \; V \left( \varphi \right) \;\; .
\end{equation}
Its pressure and its density on cosmological scales are given by
\begin{eqnarray}
\nonumber P_\varphi &=&\frac{1}{2}\dot\varphi^2 - V\left( \varphi \right)\;\;,\\
\rho_\varphi &=&\frac{1}{2}\dot\varphi^2 + V\left( \varphi \right) \;\;.
\end{eqnarray}
So, the pressure is negative if the potential dominates, and negligible if the potential equilibrates the kinetic term. Thus, a scalar field can be a good candidate for the dark fluid if it respects in particular the following constraints:\\
\indent -- its density at the time of the BBN decreases at least as fast as the density of radiation,\\
\indent -- its density from the time of last scattering to the time of structure formation evolves nearly like matter, and so $\frac{1}{2}\dot\varphi^2\approx~V\left(\varphi\right)$,\\
\indent -- after the growth of perturbations, because the scalar field is dominating the Universe, its potential does not equilibrate the kinetic term anymore and  begins to dominate; thus it will behave like a cosmological constant in the future.\\
\\
The main parameter of such a model is the same as that of quintessence models: the potential. When one considers quintessence models with real scalar field, one looks for potentials which provide a cosmological constant--like behavior today, and decreasing potentials seem to be favored. If one considers now complex scalar fields, it was shown in \cite{arbey2} that such a field can behave like cosmological matter when its potential has a dominant $m^2 |\phi|^2$ term, which corresponds to an increasing term in the potential. Thus, a way to find a ``good'' potential would be to consider a superposition of a decreasing potential which would begin to dominate today, and of the increasing quadratic term which has to dominate at least until structure formation and can nevertheless lead to an attractive effect on local scales today. A more detailed study of the possible potentials will be presented in a future publication.\\
\\
One could also hope that a scalar field might explain the excess of gravity on local scales. To do that, one can imagine a Universe filled with a scalar field. In the part -- and time -- of the Universe where the density of baryonic matter is high, the scalar field would, through gravitational interaction, have a large kinetic term which could even equilibrate the potential, so that one has an attractive net force on local scales (easier to achieve with a complex scalar field associated to an internal rotation, see \cite{arbey1}), whereas in the parts where no baryons are present, the field would not vary much, and the potential dominates, providing repulsion. Thus, on local scales, where the baryon density is high, the field behaves like matter. Where the baryonic density is small, {\it i.e.} away from galaxies and clusters, the gravitational interaction is not strong enough to increase the kinetic term of the scalar field, so that the potential dominates, and one can then observe the effects of a negative pressure. In that case, the scalar field will have a negative pressure on cosmological scales, providing a locally negative pressure in average. In the past, baryons were uniformly dense, so that the kinetic term was large everywhere, and one could have then a uniform matter behavior under these conditions. In that way, the local behavior can be in agreement with the cosmological one, and a complete cosmological scenario can be built. Such a scenario has of course to be studied further.
%%%%%%%%%%%%%%%%%%%%%%%%%%%%%%%%%%%%%%%%%%%%%%%%%%%%%%%%%%%%%%%%%%%%%%%%%%%%%%%%%%%%%%%%%%%%%%%%%%%%%%%%%%%%
%%%%%%%%%%%%%%%%%%%%%%%%%%%%%%%%%%%%%%%%%%%%%%%%%%%%%%%%%%%%%%%%%%%%%%%%%%%%%%%%%%%%%%%%%%%%%%%%%%%%%%%%%%%%
\section{Conclusion and Perspectives}
\noindent Astrophysical and cosmological observations are usually interpreted in terms of dark matter and dark energy. We have seen here that they can also be analyzed differently. Thus, it is possible to develop a model of dark fluid, which could advantageously replace a model containing in fact two dark components. Of course, hard work and studies are required to test completely the dark fluid hypothesis. Nevertheless today, as it seems difficult to find a model for dark energy and as problems concerning cold dark matter remain, it is worthwhile to investigate different ideas such as an unification of dark energy and dark matter -- that finally does not seem stranger than trying to determine the nature of two components at the same time -- which could be achieved in particular thanks to D-brane theories (in particular through the Chaplygin gas), or thanks to the so--useful scalar fields. Of course, other models may also account for dark fluid.\\
\\
An important question remains, how to interpret the dark matter problem on local scales and could the dark fluid account for the excess of gravity inside local structures? I will only provide here a qualitative analysis of whether a fluid with a negative pressure on cosmological scale can have an attractive effect on local scale, such as it is observed in galaxies (for example, with the rotation curves of spiral galaxies \cite{PSS,gentile,DDO}).\\
\\
Let us consider only the quasi--Newtonian limit of general relativity. In that case, deviations from the Minkowski metric $\eta_{\mu \nu} = {\rm diag}(1,-1,-1,-1)$ are accounted for by the perturbation $h_{\mu \nu}$. In the harmonic coordinate gauge, it satisfies the condition:
\begin{equation}
\partial_{\alpha} h^{\alpha}_{\; \mu} \, - \, \frac{1}{2} \partial_{\mu} h^{\alpha}_{\; \alpha} \; = \; 0 \;\;.
\end{equation}
One can show that the perturbation $h_{\mu \nu}$ is related to the source tensor:
\begin{equation} S_{\mu \nu} \; = \; T_{\mu \nu} \, - \, \frac{1}{2} \, g_{\mu\nu} \, T^{\alpha}_{\; \alpha}
\end{equation}
through the integral
\begin{equation}
h_{\mu \nu}\left( \vec{r} \right) \; = \; - \, 4 \, G \; {\displaystyle \int} \, {\displaystyle \frac{S_{\mu\nu}\left(\vec{r}\,'\right)}{|\,\vec{r}\,'\,-\,\vec{r}\,|}}\,d^3\vec{r}\,' \;\; .
\end{equation}
If the energy--momentum tensor is written as:
\begin{equation}
T^{\mu\nu}=(P+\rho) U^\mu U^\nu - P g^{\mu\nu}\;\;,
\end{equation}
one can show that, at leading order in $h_{\mu \nu}$,
\begin{equation}
S^{\mu\nu}=(P+\rho) U^\mu U^\nu - \frac{1}{2} \eta^{\mu\nu} (\rho-P) \;\;.
\end{equation}
The gravitational potential is in fact $\Phi = h_{00}/2$, so that:
\begin{equation}
\Phi\left( \vec{r} \right) \; = \; - \, 2 \, G \; {\displaystyle \int} \, {\displaystyle \frac{S_{00}\left(\vec{r}\,'\right)}{|\,\vec{r}\,'\,-\,\vec{r}\,|}}\,d^3\vec{r}\,' \;\; .
\end{equation}
For a fluid at rest, $U^\mu=(1,0,0,0)$, with an equation of state $P=\omega \rho$, one has:
\begin{equation}
S_{00}=\frac{1}{2} \rho (1+3 \omega) \;\;,
\end{equation}
and consequently this fluid has an attractive effect only if $\omega > -1/3$. From the study of supernov\ae~it seems that our dark fluid is not in that state today, so that its effects are mainly repulsive. Nevertheless, it is possible that the dark fluid has a different behavior on cosmological and on local scales. Indeed, the density and pressure of the fluid on cosmological scale are spacial averages of the local density and pressure, and one can assume that:
\begin{eqnarray}
\nonumber \rho\left(t,\vec{r}\right) &=& \rho^{\mbox{cosmo}}\left(t\right) + \delta\rho\left(t,\vec{r}\right)\;\;,\\
P\left(t,\vec{r}\right) &=& P^{\mbox{cosmo}}\left(t\right) + \delta P\left(t,\vec{r}\right) \;\;,
\end{eqnarray}
where $\rho^{\mbox{cosmo}}$ and $P^{\mbox{cosmo}}$ are the cosmological density and pressure, with the spatial averages:
\begin{equation}
<\delta\rho\left(t,\vec{r}\right)>=<\delta P\left(t,\vec{r}\right)>= 0 \;\;.
\end{equation}
The density of dark fluid on cosmological scales today is of the order of the critical density, i.e. $\rho^0_c\approx~9\times~10^{-29}\mbox{g.cm}^{-3}$. One can compare it to the estimated matter density in the Milky Way at the radius of the Sun $\rho^{\mbox{Sun}} \approx 5 \times 10^{-24}\mbox{g.cm}^{-3}$ \cite{milky_way}. Hence, even if the dark fluid's local density would represent 1\% of this total matter local density only, its value would be much higher than the cosmological densities today. Therefore, on local scales, one can assume that $\delta\rho\left(t,\vec{r}\right) \gg \rho^{\mbox{cosmo}}\left(t\right)$ and thus write:
\begin{equation}
S_{00} \approx \frac{1}{2} (\delta\rho + 3 \delta P) \;\;.
\end{equation}
To have a net attraction, we get finally the same kind of constraint as before, $\delta\rho > -3 \delta P$, but this time we do not have to use the cosmological constraints, because the local behavior of the dark fluid can be very different from the cosmological one. Moreover, if no model is specified, one can still hope that $\delta P$ could be negligible on local scales, so that the local behavior of the dark fluid is matter-like, and that the usual Newtonian equation can be retrieved:
\begin{equation}
\Phi\left( \vec{r} \right) \; = \; - \, G \; {\displaystyle \int} \, {\displaystyle \frac{\delta\rho\left(\vec{r}\,'\right)}{|\,\vec{r}\,'\,-\,\vec{r}\,|}}\,d^3\vec{r}\,' \;\; .
\end{equation}
This local behavior will of course have to be verified quantitatively for each dark fluid model, but nevertheless gives hope for a unified explanation on any scale. In particular, considering scalar fields, this qualitative analysis tends to show that it would be interesting to try to find a potential which gives today a negative pressure on cosmological scale, but which also gives a matter behavior in local structures, {\it i.e.} where the density of baryons is high.\\
To conclude, the dark fluid appears as an interesting possibility to explain the observations. As the properties of the dark fluid are different from dark matter and dark energy, models of dark fluid are worth to be studied, and we can now use many precise observations as strong constraints on such models.
%%%%%%%%%%%%%%%%%%%%%%%%%%%%%%%%%%%%%%%%%%%%%%%%%%%%%%%%%%%%%%%%%%%%%%%%%%%%%%%%%%%%%%%%%%%%%%%%%%%%%%%%%%%%%
%%%%%%%%%%%%%%%%%%%%%%%%%%%%%%%%%%%%%%%%%%%%%%%%%%%%%%%%%%%%%%%%%%%%%%%%%%%%%%%%%%%%%%%%%%%%%%%%%%%%%%%%%%%%%%
\section*{Acknowledgements}
\noindent I would like to thank Farvah Mahmoudi, H\'el\`ene Courtois, Julien Devriendt, Thierry Sousbie and Wolfgang Hillebrandt for their comments and for useful discussions.
%%%%%%%%%%%%%%%%%%%%%%%%%%%%%%%%%%%%%%%%%%%%%%%
%                      BIBLIOGRAPHY
%%%%%%%%%%%%%%%%%%%%%%%%%%%%%%%%%%%%%%%%%%%%%%%

%%%%%%%%%%%%%%%%%%%%%%%%%%%%%%%%%%%%%%%%%%%%%%%
%%%%%%%%%%%%%%%%%%%%%%%%%%%%%%%%%%%%%%%%%%%%%%%
\end{document}